\documentclass[twocolumn,showpacs,prl]{revtex4}

\usepackage{amsmath,amssymb}
\usepackage{graphicx}
\usepackage{verbatim}

\draft
\begin{document}

\title{Optically Induced Spin Hall Effect in Atoms}

\author{Xiong-Jun Liu$^{a}$\footnote{Electronic address:
phylx@nus.edu.sg}, Xin Liu$^{b,c}$, L. C. Kwek$^{a,d}$ and C. H.
Oh$^{a}$\footnote{Electronic address: phyohch@nus.edu.sg}}
\affiliation{a. Department of Physics, National University of
Singapore,
2 Science Drive 3, Singapore 117542 \\
b. Department of Physics, Texas A\&M University, College Station,
Texas 77843-4242, USA\\
c. Theoretical Physics Division, Chern Institute of
Mathematics, Tianjin 300071, P.R.China\\
d. National Institute of Education, Nanyang Technological
University, 1 Nanyang Walk, Singapore 639798}

\begin{abstract}
We propose an optical means to realize a spin hall effect (SHE) in
neutral atomic system by coupling the internal spin states of
atoms to radiation. The interaction between the external optical
fields and the atoms creates effective magnetic fields that act in
opposite directions on ``electrically" neutral atoms with opposite
spin polarizations. This effect leads to a Landau level structure
for each spin orientation in direct analogy with the familiar SHE
in semiconductors. The conservation and topological properties of
the spin current, and the creation of a pure spin current are
discussed.
\end{abstract}
\date{\today }
\maketitle

\indent Information devices based on spin states of particles
require a lot less power consumption than equivalent charge based
devices \cite{spintronics1}. To implement practical spin-based
logical operations, a basic underlying theory, i.e. spin hall
effect (SHE) has been widely studied for the creation of spin
currents in semiconductors \cite{zhang,niu,hall1,hall2}. Nearly
all current publications on SHE involve some form of spin-orbit
coupling, including the interaction between charged particles in
semiconductors and external electric field. The physics of SHE in
semiconductors is: in the presence of spin-orbit coupling, the
applied electric field leads to a transverse motion (perpendicular
to the electric field), with spin-up and spin-down carriers moving
oppositely to each other, creating a transverse spin current.
However, spin current can also be generated by interacting optical
fields with charged particles in semiconductors
\cite{optical1,optical2}, even in absence of spin-orbit coupling
\cite{liu}.

In this letter, we show how SHE can be induced by optical fields
in neutral atomic system. Quantum states of atoms can be
manipulated by coupling their internal degrees of freedom (atomic
spin states) to radiation, making it possible to control  atomic
spin propagation through optical methods. We consider here an
ensemble of cold Fermi atoms interacting with two external light
fields (Fig. 1). The ground ($|g_\pm,\pm\frac{1}{2}\rangle$) and
excited ($|e_\pm,\pm\frac{1}{2}\rangle$) states are hyperfine
angular momentum states (atomic spins) with their total angular
momenta $F_g=F_e=1/2$.
\begin{figure}[ht]
\includegraphics[width=0.7\columnwidth]{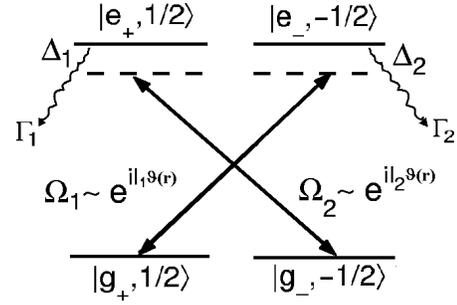}
\caption{Fermi atoms with four-level internal hyperfine spin
states interacting with two light fields. This can be
experimentally realized with alkali atoms, such as $^6$Li atoms
($2^2S_{1/2}(F=1/2)\longleftrightarrow2^2P_{1/2}(F'=1/2)$)
\cite{li}.} \label{fig1}\end{figure} The transitions from
$|g_-,-\frac{1}{2}\rangle$ to $|e_+,\frac{1}{2}\rangle$ and from
$|g_+,\frac{1}{2}\rangle$ to $|e_-,-\frac{1}{2}\rangle$ are
coupled respectively by a $\sigma_+$ light with the Rabi-frequency
$\Omega_2=\Omega_{20}\exp(i(\bold k_2\cdot\bold r+l_2\vartheta))$
and by a $\sigma_-$ light with the Rabi-frequency
$\Omega_1=\Omega_{10}\exp(i(\bold k_1\cdot\bold r+l_1\vartheta))$,
where $\bold k_{1,2}=k_{1,2}\bold{\hat e}_z$ are the wave-vectors
and $\vartheta=\tan^{-1}({y}/{x})$. $l_1$ and $l_2$ indicate that
$\sigma_+$ and $\sigma_-$ photons are assumed to have the orbital
angular momentum $\hbar l_1$ and $\hbar l_2$ along the $+z$
direction, respectively \cite{angular}. For simplicity, we replace
the notations $|\alpha_\pm,\pm\frac{1}{2}\rangle$ by
$|\alpha_\pm\rangle \ (\alpha=e,g)$. The $\bold r$-representation
atomic wave function is denoted by $\Phi_\alpha(\bold r,t)$. It is
helpful to introduce the slowly-varying amplitudes of atomic
wave-functions by (setting $\omega_{g_\pm}=0$):
$\Psi_{g_\pm}=\Phi_{g_\pm}, \Psi_{e_+}=\Phi_{e_+}(\bold
r,t)e^{-i(\bold k_2\cdot\bold r-(\omega_{e_+}-\Delta_1)t)},
\Psi_{e_-}=\Phi_{e_-}(\bold r,t)e^{-i(\bold k_1\cdot\bold
r-(\omega_{e_-}-\Delta_2)t)}$, where $\hbar\omega_\alpha$ is the
energy of the state $|\alpha\rangle$, $\Delta_{1,2}$ are
corresponding detunings. The total Hamiltonian $H=H_0+H_1+H_2$ of
the system reads:
\begin{eqnarray}\label{eqn:Hamiltonian1a}
H_0&=&\sum_{\alpha=e_{\pm}g_{\pm}}\int
d^3r\Psi_{\alpha}^*\bigr(-\frac{\hbar^2}{2m}\nabla^2+V(\bold
r)\bigr) \Psi_{\alpha},\nonumber\\
H_1&=&\hbar\Delta_1\int
d^3r\Psi^*_{e_+}S_{e_+e_+}\Psi_{e_+}\nonumber\\
&&+\hbar\int
d^3r(\Psi^*_{e_+}\Omega_{10}e^{il_1\vartheta}S_{1+}\Psi_{g_-}+h.a.),\\
H_2&=&\hbar\Delta_2\int d^3r\Psi^*_{e_-}S_{e_-e_-}\Psi_{e_-}\nonumber\\
&&+\hbar\int
d^3r(\Psi^*_{e_-}\Omega_{20}e^{il_2\vartheta}S_{2+}\Psi_{g_+}+h.a.),\nonumber
\end{eqnarray}
where the atomic operators
$S_{e_{\pm}e_{\pm}}=|e_{\pm}\rangle\langle e_{\pm}|$,
$S_{1+}=|e_+\rangle\langle g_-|, S_{2+}=|e_-\rangle\langle g_+|$,
$S^{\dag}_{1+}=S_{1-}$, $S^{\dag}_{2+}=S_{2-}$, and $V(\bold r)$
is an external trap potential. The collisions (s-wave scattering)
between cold Fermi atoms are negligible. The interaction part of
the Hamiltonian can be diagonalized with a local unitary
transformation: $\tilde{H}_I=U(\bold r)H_IU^{\dag}(\bold
r)=U(\bold r)(H_1+H_2)U^{\dag}(\bold r)$ where
\begin{eqnarray}\label{eqn:unit1}
U(\bold r)={\left[ \begin{matrix} U_1& 0\\
0& U_2\end{matrix} \right]}
\end{eqnarray}
with
\begin{eqnarray}\label{eqn:unit2}
U_j={\left[ \begin{matrix} \cos\theta_j& \sin\theta_je^{il_j\vartheta(\bold r)}\\
-\sin\theta_je^{-il_j\vartheta(\bold r)}& \cos\theta_j\end{matrix}
\right]}, \ \ j=1,2.\nonumber
\end{eqnarray}
Under this transformation the four eigenstates of interaction
Hamiltonian can be obtained as $\bigr[|\psi_+\rangle,
|\psi_-\rangle\bigr]^{{\small T}}=U_1\bigr[|e_+\rangle,
|g_-\rangle\bigr]^{T}$ and $\bigr[|\phi_+\rangle,
|\phi_-\rangle\bigr]^{\small T}=U_2\bigr[|e_-\rangle,
|g_+\rangle\bigr]^{T}$. The mixing angles $\theta_j$ are defined
by $\tan\theta_{1,2}=E_{\psi,\phi}^-/|\Omega_{1,2}|$, where the
eigenvalues of $|\psi_{\pm}\rangle$ and $|\phi_{\pm}\rangle$ can
be calculated by:
$E_{\psi,\phi}^{\pm}=(\Delta_{1,2}\pm\sqrt{\Delta_1^2+4|\Omega_{1,2}|^2})/2$.

In order to suppress the spontaneous emission of the excited
states, we consider the large detuning case, i.e.
$\Delta_j^2\gg\Omega_{j0}^2$. By calculating all values up to the
order of $\Omega_{j0}^2/\Delta_j^2$, one can verify that
$\tan^2\theta_j=\Omega^2_{j0}/(\Delta_j^2-\Omega_{j0}^2)$, and
$E^+_{\psi,\phi}\gg E^-_{\psi,\phi}$. We then invoke the adiabatic
condition so that the population of the higher levels
$|\psi_+\rangle$ and $|\phi_+\rangle$ is adiabatically eliminated.
Moreover, the total system is confined to the ground eigenstates
\begin{eqnarray}\label{eqn:state1}
|\Psi\rangle=\cos\gamma|S_\downarrow\rangle+\sin\gamma|S_\uparrow\rangle,
\end{eqnarray}
where to facilitate further discussions, we have put the effective
spin states \cite{liu}: $|S_\downarrow\rangle=|\psi_-\rangle$,
$|S_\uparrow\rangle=|\phi_-\rangle$ with their $z$-component
effective spin polarizations $S_z^{\uparrow}\approx\hbar/2$ and
$S_z^{\downarrow}\approx-\hbar/2$. The parameter $\gamma$
describes the probability of an atom in states
$|S_\downarrow\rangle$ and $|S_\uparrow\rangle$, determined
through the initial condition. One can also see the population of
excited states $|e_\pm\rangle$ is very small, therefore the atomic
decay can be neglected in the present situation.

Under adiabatic condition, the local transformation $U(\bold r)$
leads to a diagonalized $SU(2)$ gauge potential: $(e/c)\bold
A_\alpha=-i\hbar\langle S_\alpha|\bold\nabla|S_\alpha\rangle, \
(\alpha=\downarrow,\uparrow)$. Accordingly, the effective (scalar)
trap potentials read $V_\alpha(\bold r)=V(\bold
r)-\hbar|\Delta_j|^{-1}\Omega^2_{j0}-(2m)^{-1}|\hbar\langle
S_\alpha|\bold\nabla|S_\alpha\rangle|^2$ with $j=1$ (for
$\alpha=\downarrow$) and $j=2$ (for $\alpha=\uparrow$). Specially
we have $\bold A_\downarrow=-\bold A_\uparrow=\hbar
lce^{-1}\frac{\Omega_0^2}{\Delta^2}(x\hat e_y-y\hat e_x)/\rho^2$
and $V_{eff}(\bold r)=V_{\uparrow,\downarrow}(\bold r)=V(\bold
r)-\hbar\Omega^2_{0}/\Delta-\hbar^2l^2\Omega_0^4/(2m\Delta^4\rho^2)$
with $\rho=\sqrt{x^2+y^2}$, when we choose
$\Delta_1=\Delta_2=\Delta$, $\Omega_{10}=\Omega_{20}=\Omega_0$ and
$l_1=-l_2=l$, i.e. the angular momenta of the two light fields are
opposite in direction. This result is intrinsically interesting:
by coupling the atomic spin states to radiation, we find the
atomic system can be described as an ensemble of charged particles
with opposite spins experiencing magnetic fields in opposite
directions but subject to the same electric field. With the help
of gauge and trap potentials, we rewrite the Hamiltonian
effectively as
\begin{eqnarray}\label{eqn:Hamiltonian2}
H&=&\int
d^3r\Psi_{s_\downarrow}^{*}\bigr[\frac{1}{2m}(\hbar\partial_k+i\frac{e}{c}A_k)^2\bigr]
\Psi_{s_\downarrow}\nonumber\\
&&+\int
d^3r\Psi_{s_\uparrow}^{*}\bigr[\frac{1}{2m}(\hbar\partial_k-i\frac{e}{c}A_k)^2\bigr]
\Psi_{s_\uparrow}\\&&+\int
d^3r(V_\downarrow|\Psi_{s_\downarrow}|^2+V_\uparrow|\Psi_{s_\uparrow}|^2)),\nonumber
\end{eqnarray}
where $\Psi_{s_\downarrow}(\bold r)=\cos\gamma\langle\bold
r|S_\downarrow\rangle$ and $\Psi_{s_\uparrow}(\bold
r)=\sin\gamma\langle\bold r|S_\uparrow\rangle$ are spin wave
functions in $\bold r$-representation.

Before calculating the spin current, we study the properties of
spin currents in our model. We first consider a conservation law.
The general spin density in the present system is calculated using
$\vec S(\bold r,t)=\Psi_{s_\downarrow}^{*}\vec
S_\downarrow\Psi_{s_\downarrow}+\Psi_{s_\uparrow}^*\vec
S_\uparrow\Psi_{s_\uparrow}$ with $\vec
S_{\uparrow\downarrow}=(S_x^{\uparrow\downarrow},S_y^{\uparrow\downarrow},S_z^{\uparrow\downarrow})$.
Moreover, the spin current density $\vec J_k(\bold r,
t)=(J_k^{s_x},J_k^{s_y},J_k^{s_z})$ is defined by
\begin{eqnarray}\label{eqn:spincurrent1}
\vec J_k&=&-\frac{i\hbar}{m}\vec S_{\downarrow}(\Psi_{s_\downarrow}^*D_{1k}\Psi_{s_\downarrow}-\Psi_{s_\downarrow}D^*_{1k}\Psi_{s_\downarrow}^*)\nonumber\\
&&-\frac{i\hbar}{m}\vec
S_{\uparrow}(\Psi_{s_\uparrow}^*D_{2k}\Psi_{s_\uparrow}-\Psi_{s_\uparrow}D^*_{2k}\Psi_{s_\uparrow}^*),
\end{eqnarray}
where $D_{1k}=\partial_k+i\frac{e}{c}A_{k}$ and
$D_{2k}=\partial_k-i\frac{e}{c}A_{k}$ are the covariant derivative
operators. By a straightforward calculation, we verify the
following continuity equation
\begin{eqnarray}\label{eqn:continuity}
\partial_t\vec S(\bold r,t)+\partial_k\vec J_k=A_k\sigma_z\hat{\bold
e}_z\times\vec J_k(\bold r, t),
\end{eqnarray}
where $\sigma_z$ is the usual Pauli matrix. It is easy to see
that  the right hand side of Eq. (\ref{eqn:continuity}) equals
zero for the $s_z$-component spin current. Thus the spin current
$J_k^{s_z}$ is conserved. But there is no spin-orbit coupling in
the present atomic system. Thus, it is easily verified that the
orbit-angular momentum current is also conserved. These results
underlines the conservation law for $s_z$-component of total
angular momentum current of atoms. Secondly, it is interesting
that the two spin functions in the effective Hamiltonian
(\ref{eqn:Hamiltonian2}) are not independent: there is a
nontrivial effective coupling mediated by an effective
electromagnetic field. This implies that the spin current in our
model could have a nontrivial hidden topology. We next discuss
the topological properties of $J_k^{s_z}$. We note that
$\Psi_{s_\alpha}=n_a^{1/2}\zeta_\alpha$, where the complex
$\zeta_\alpha=|\zeta_\alpha|e^{-i\varphi_\alpha}$ with
$|\zeta_\downarrow|^2+|\zeta_\uparrow|^2=1$, and $n_a$ is the
total atomic density which is assumed to be constant in our case.
The $s_z$-component of eq. (\ref{eqn:spincurrent1}) can be recast
as
\begin{eqnarray}\label{eqn:spincurrent2}
J_k^{s_z}&=&-i\frac{n_a\hbar^2}{2m}(\zeta_\downarrow\partial_k\zeta_\downarrow^*-
\zeta_\downarrow^*\partial_k\zeta_\downarrow+\zeta_\uparrow^*\partial_k\zeta_\uparrow-
\zeta_\uparrow\partial_k\zeta_\uparrow^*)-\nonumber\\
&&-\frac{n_a\hbar^2e}{mc}A_k.
\end{eqnarray}
Furthermore, we introduce the unit vector field,
$\vec\lambda=(\bar\zeta, {\vec\sigma}\zeta)$, where
$\zeta=(\zeta_\downarrow,\zeta_\uparrow^*)^T$ and
$\vec\sigma=(\sigma_x,\sigma_y,\sigma_z)$. It then follows that
$\lambda_1=\zeta_\downarrow^*\zeta_\uparrow^*+\zeta_\downarrow\zeta_\uparrow,
\lambda_2=i(\zeta_\downarrow\zeta_\uparrow-\zeta_\downarrow^*\zeta_\uparrow^*)$
and $\lambda_3=|\zeta_\downarrow|^2-|\zeta_\uparrow|^2$. Using
these definitions we have
\begin{eqnarray}\label{eqn:spincurrent3}
(\bold\nabla\times{\bold
J}^{s_z})_m=-\frac{n_a\hbar^2e}{mc}B_m-\frac{n_a\hbar^2}{2m}\epsilon_{mkl}\vec\lambda
\cdot(\partial_k\vec\lambda\times\partial_l\vec\lambda).
\end{eqnarray}
The contribution
$\vec\lambda\cdot(\partial_k\vec\lambda\times\partial_l\vec\lambda)$
provides a topological term in the spin current induced by optical
fields. Note each value of $\vec\lambda$ represents a point in the
two-dimensional sphere $S^2$. The variation of $\vec\lambda$
depends on the relative density of the two spin components
$\gamma(\bold r)$ and the sum of the phase spreadings
$\varphi_\uparrow(\bold r)+\varphi_\downarrow(\bold r)$. It is
easily verified that $\vec\lambda$ can cover the entire surface
$S^2$, when the parameter distributions in the interaction region
satisfy: $\varphi_1+\varphi_2: 0\rightarrow2n_1\pi$ and $\gamma:
0\rightarrow n_2\pi/2$ with integers $n_1,n_2\geq1$. We then
obtain a map degree between the unit vector field $\vec\lambda$
and the spatial vectors $F:\vec\lambda\rightarrow\bold r/r$ so
that the closed-surface integral
$$\oint_s(\bold\nabla\times{\bold J}^{s_z})\cdot d\bold
S\sim\oint_sds_m\epsilon_{mkl}\frac{\bold
r}{r}\cdot(\partial_k\frac{\bold r}{r}\times\partial_l\frac{\bold
r}{r})\sim4\pi n,$$ where $n$ is the winding number. In physics,
such mapping corresponds to the formation of a local inhomogeneity
in the densities of the spin-up and spin-down atoms. The
Hamiltonian (\ref{eqn:Hamiltonian2}) can also be regarded as a
system of $two$-$flavor$ oppositely charged particles interacting
with $one$ external effective magnetic field. Such model has a
wide range of applications to, e.g. two-band superconductivity
\cite{super}, etc. It is interesting to note that, under certain
conditions liquid metallic hydrogen, as an example, might allow
for the coexistence of superconductivity with both electronic and
protonic Cooper pairs \cite{cooper}. Faddeev et al. have also
discovered a series of nontrivial topological properties in such
systems \cite{faddeev}. Based on this technique, we have developed
an optical way to realize $two$-$flavor$ artificially charged
system, which could provide a deeper understanding of the
essential physical mechanisms in cold atomic systems.

For practical application, an important goal is to create a pure
spin current injection, where the massive current is zero. To this
end, we propose using a columnar spreading light fields that
\cite{angular} $\Omega_{0}(\bold r)=f\rho$ with $f>0$.
Furthermore, we set the $x$-$y$ harmonic trap $V(\bold
r)=\frac{1}{2}m\omega^2_\perp(\vec\rho+x_0\bold{\hat e}_x)^2$
centered at $\vec\rho=-x_0\bold{\hat e}_x$, where the frequency is
tuned to $\omega_\perp^2=(1+\frac{\hbar
l^2f^2}{2m\Delta^3})\frac{2\hbar f^2}{m\Delta}$. It should be
emphasized that the (positive) potential $V(\bold r)$ can
typically be obtained with a blue-detuned dipole trap method
\cite{trap} for instance. The uniform magnetic and electric fields
corresponding to the gauge and scalar potentials are
\begin{eqnarray}\label{eqn:field1} \bold B_\downarrow(-\bold
B_\uparrow)=\frac{\hbar l c}{e}\frac{f^2}{\Delta^2}\hat e_z,
\bold E=-\bigr(1+\frac{\hbar
l^2f^2}{4m\Delta^3}\bigr)\frac{2\hbar f^2x_0}{e\Delta}\hat e_x,
\end{eqnarray}
along the $z$ ($-z$) and $x$ directions respectively. Atoms in
different spin states $|S_\alpha\rangle$ experience the opposite
magnetic fields $\bold B_\alpha$ but the same electric field
$\bold E$. This leads to a Landau level structure for each spin
orientation. To calculate the spin currents explicitly, one needs
to obtain the eigenstates of the present system. However, the
present forms of gauge and scalar potentials in the Hamiltonian
(\ref{eqn:Hamiltonian2}) cannot be diagonalized easily;
consequently, we need to resort to perturbation theory to
calculate the spin currents. The Hamiltonian for a single atom in
state $|S_{\alpha}\rangle$ can be written as:
$H^{\alpha}=H^{\alpha}_0+H'$ ($\alpha=\downarrow,\uparrow$), where
the perturbation part $H'=-eEx$ and
\begin{eqnarray}\label{eqn:hamiltonian3}
H^{\alpha}_0=\frac{\hbar^2eB}{2mc}(R_{\alpha}^2+P_{\alpha}^2)+\frac{1}{2m}p_z^2
\end{eqnarray}
with
$R_{\uparrow\downarrow}=(\frac{c}{eB})^{1/2}(p_x-\frac{e}{c}A^{\uparrow\downarrow}_x),
P_{\uparrow\downarrow}=(\frac{c}{eB})^{1/2}(p_y-\frac{e}{c}A^{\uparrow\downarrow}_y)$
and $B=|\bold B_{\uparrow,\downarrow}|$. One can verify that
$[R_{\alpha},P_{\beta}]=i\hbar\delta_{\alpha\beta}$, so the
eigenfunction $\mu^{\alpha}_{nk}(R)$ of $H^{\alpha}_0$ is Hermite
polynomial with the Landau level
$E_{nk,\alpha}=(n+1/2)\hbar\omega+\hbar k^2/2m$ and
$\omega=eB/mc$. For the weak field case, the spin/massive current
carried by an atom can be calculated perturbatively  to the first
order correction on the state $|\mu^{\alpha}_{nk}\rangle$
\begin{eqnarray}\label{eqn:spincurrent4}
(j^y_{s_z,m})_{nk,\alpha}&=&\langle\mu_{nk}^\alpha|j^y_{s_z,m}|\mu_{nk}^{\alpha}\rangle+\nonumber\\
&&+\bigl(\sum_{n'\alpha'}\frac{\langle\mu^{\alpha'}_{n'k}|H'|\mu^{\alpha}_{nk}
\rangle\langle\mu^{\alpha}_{nk}|j^y_{s_z,m}|\mu^{\alpha'}_{n'k}\rangle}{E_{n,\alpha}-E_{n',\alpha'}}\nonumber\\
&&+h.a.\bigl),
\end{eqnarray}
where the spin current operator is
$(j^y_{s_z})_{\alpha}=\frac{\hbar}{2}(S^{\alpha}_zv^{\alpha}_y+v^{\alpha}_yS^{\alpha}_z)$
with $v^{\alpha}_y=[y,H^{\alpha}]/ih=(eB/m^2c)^{1/2}P_{\alpha}$
and the massive current operator $j_{m}=mv_y$. It is easy to see
that
$\langle\mu^{\alpha}_{nk}|j^y_{s_z,m}|\mu^{\alpha'}_{n'k}\rangle=0$
when $n'\neq n\pm1$, thus only the terms with $n'=n\pm1$
contribute in the above equation.

If the spatial spreads of the atoms in $x$ and $z$ directions are
$L_x$ and $L_z$, respectively, the average current density for the
total system is given by $J^y_{s_z,m}=\frac{1}{L_xL_z}\int
dE\bigr((j^{(0)}_{s_z,m})_{nk,\alpha}
+(j^{(1)}_{s_z,m})_{nk,\alpha}\bigr)f(E)$ with $f(E)$ as the Fermi
distribution function. For $^6$Li atoms we may consider the
initial condition that $\sin^2\gamma=\cos^2\gamma=1/2$, i.e. the
atoms have the equal probability in state $|S_\downarrow\rangle$
and $|S_\uparrow\rangle$, as in the usual optical trap
\cite{trap1}. Rewriting the perturbation part as $H'=i\hbar
E(\frac{ec}{B})^{1/2}\frac{\partial}{\partial R}$ and substituting
this result into Eq (\ref{eqn:spincurrent4}) and finally we get
\begin{eqnarray}\label{eqn:spincurrent7}
J^y_{s_z}=n_a(\hbar+\frac{\hbar^2l^2f^2}{4m\Delta^3})\frac{\Delta
x_0}{l},
\end{eqnarray}
whereas the massive current $J_m$ is zero. In fact, under present
interaction of external effective electric and magnetic fields,
the atoms with opposite velocities in $y$ direction have opposite
spin polarizations (see Fig.2). Thus the massive current vanishes
whereas a pure spin current is obtained. This result allows us to
create conserved spin currents without using atomic beams.
\begin{figure}[ht]
\includegraphics[width=0.8\columnwidth]{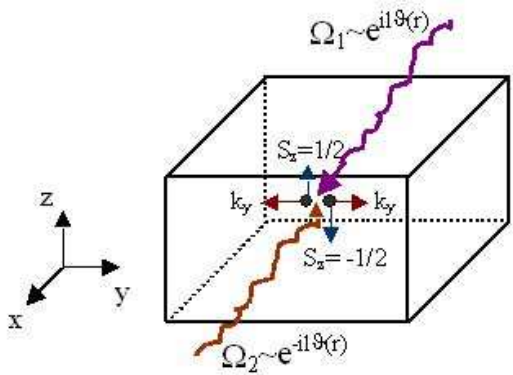}
\caption{(Color online) Under the interaction of effective
electric and magnetic fields induced by light fields, atoms in
state $|S_\downarrow\rangle$ and $|S_\uparrow\rangle$ have
opposite momenta in $y$ direction.} \label{fig1}\end{figure}

To observe SHE in present cold atomic system, a spin-sensitive
measurement \cite{niu1} can be used. For example, the technique of
atom chip \cite{atomchip} can be used to implement the spin
current described in the present model. The created spin current
can lead to spin accumulations on opposite sides along $y$
direction of the atom chip. Experimentally, one could detect such
spatially separated spin polarizations using magnetic resonance
force microscopy (MRFM) \cite {mrfm} for instance. Another means
to detect the spin accumulation can also be achieved with
fluorescence. For $^6$Li atoms, one may perform resonant Raman
transitions from the accumulated spin state $S_z=-1/2$ to
$2^2P_{1/2} (F=3/2, m=-3/2)$ or from $S_z=1/2$ to $2^2P_{1/2}
(F=3/2, m=3/2)$ ($D_1$ line) and then observe the fluorescence. As
long as the spin accumulations are separated to say larger than
$5-10$ microns, the two separated images can be resolved in an
experiment.

One should also discuss the adiabatic condition employed in above
calculations. Atomic motion may lead to transitions between ground
eigenstates and excited ones, e.g. the element of transition
between $|\psi_+\rangle$ and $|\psi_-\rangle$ can be calculated by
$\tau_\pm=|\langle\psi_+|\partial_t|\psi_-\rangle|=|\bar{\bold
v}\cdot\nabla\theta_1(\bold
r)+\frac{1}{2}l_1\sin2\theta_1\bar{\bold
v}\cdot\nabla\vartheta(\bold r)|$ where $\bar{\bold v}$ is the
average resulting velocity in spin currents. The adiabatic
condition requires $\tau_\pm\ll|E_\psi^+-E_\psi^-|$. For a
numerical evaluation one typically set the parameters
$\Delta\sim10^8$s$^{-1}$, $l<10^4, f=5\times10^{10}$(s$\cdot$
m)$^{-1}$, $x_0 \approx 2.0 \mu m$. We then find the velocity
$\bar{v}<1.0$ m/s and
$\tau_\pm/|E_\psi^+-E_\psi^-|\sim10^{-3}\ll1$, which guarantees
the validity of the adiabatic condition. In the case of $^6$Li
atomic system, the $x$-$y$ trap potential $V(\bold r)$ can be
achieved through $D_2$ transition (from $2^2S_{1/2}$ to
$2^2P_{3/2}$) with a blue detuning \cite{li,trap}. With former
parameters and by tuning the trap frequency to
$\omega_\perp\approx 728.5$Hz, one achieves the uniform magnetic
and electric fields in Eq. (\ref{eqn:field1}). Furthermore, if we
employ an atomic system with the atomic density $n_a \approx1.0
\times10^{10}$cm$^{-3}$, one can find the spin current
$J^y_{s_z}\approx 1.322 \times10^{-5}$eV/cm$^{2}$. On the other
hand, we note that for all practical purposes, the light fields
have a finite cross section $S_{xy}$, which may have a boundary
effect on the calculation of spin currents. For the cold Fermi
atomic gas, the spatial scale of the interaction region is about
$0.1$ mm \cite{trap1}, then such boundary effect can be neglected
since $lf^2S_{xy}/\Delta^2\sim10^2\gg 1$. This means the effective
magnetic flux induced by the light fields can support a
sufficiently large degeneracy at each Landau level.

In conclusion we proposed an optical means to realize a new type
of spin hall effect in the neutral atomic system. The spin current
created in this way is conserved and could possess interesting
topological properties. The present atomic system is equivalent to
a {\it two-flavor} artificially charged system, providing a direct
analogy between the dynamics of electrons in solid systems, e.g.
two-band superconductors \cite{super,cooper}, and the behavior of
cold atoms in optical potentials. The effect also provides
understanding of the basic physical mechanisms of SHE with a wide
range of applications in cold atomic systems.

We thank M. Barrett, R. Ballagh and J. Fuchs for fruitful
discussions and helpful comments. This work is supported by NUS
academic research Grant No. WBS: R-144-000-172-101, US NSF under
the grant DMR-0547875, and NSF of China under grants No.10275036.



\end{document}